\documentclass[twocolumn,prl,showpacs,byrevtex,floatfix]{revtex4-1}
\usepackage{epsf,graphicx,amssymb}
\usepackage[usenames]{color}

\begin{document}

\title{Dissipated power within a turbulent flow\\ forced homogeneously by magnetic particles}

\author{Eric Falcon}
\email[E-mail: ]{eric.falcon@univ-paris-diderot.fr}
\author{Jean-Claude Bacri}
\author{Claude Laroche}
\affiliation{Univ Paris Diderot, Sorbonne Paris Cit\'e, MSC, UMR 7057 CNRS, F-75 013 Paris, France}

\date{\today}

\begin{abstract}
We report measurements of global dissipated power within a turbulent flow homogeneously forced at small scale by a new forcing technique. The forcing is random in both time and space within the fluid by using magnetic particles in an alternating magnetic field. By measuring the growth rate of the fluid temperature, we show how the dissipated power is governed by the external control parameters (magnetic field, and number $N$ of particles). We experimentally found that the mean dissipated power scales linearly with these parameters, as expected from the magnetic injected power scalings. These experimental results are well described by simple scaling arguments showing that the main origins of the energy dissipation are due to viscous turbulent friction of particles within the fluid and to the inelasticity of collisions. Finally, by measuring the particle collision statistics, we also show that the particle velocity is independent of $N$, and is only fixed by the magnetic ``thermostat''. 
\end{abstract}


    
\maketitle

\section{Introduction}
When a magnetic fluid (i.e. nanometric ferromagnetic particles within a carrier liquid) is subjected to an alternative magnetic field, an intense heat is released at the particle scale that may serve for magnetic hyperthermia in medical therapies \cite{Lesfilles}. In this context, many studies have been performed to quantify the heat production within a ferrofluid subjected to an alternating magnetic field \cite{Rosensweig02,Skumiel2013}. 

In soft matter, diluted granular media made of permanent magnetic particles of centimetric scale, subjected to an alternating magnetic field, have been used in 2D \cite{Bologa1982,Schmick08} or 3D \cite{FalconEPL13} experiments to test dissipative granular gas theories. This type of forcing has been also used to study pattern formation of magnetic particles in suspensions on liquid surface \cite{Snezhko05}, and to study the mechanical energy conversion into heat for a granular system within a gas or a liquid \cite{Buevich1985,Bologa1986}.

For closed turbulent flows, most laboratory experiments on 3D turbulence uses a spatially localized forcing, at large scale, in a deterministic way (e.g., oscillating grids; rotating blades). Here, we present a new forcing technique where the fluid is forced randomly in both time and space, at small scale, using magnetic particles. It allows a forcing within the bulk, which favors the homogeneity of the fluid velocity field.

Our magnetic particle consists in a small permanent magnet encapsulated in a soft shell of centimetric size. The torque applied by the alternating magnetic field over the magnetic moment of each particle leads to a random forcing, in both space and time, where particles are driven by chaotic rotational motions. The collisions between particles or particles/walls lead to translational erratic motions of the particles. We have previously used such particles to experimentally study a 3D diluted granular medium in air driven randomly in both time and space \cite{FalconEPL13}. This homogeneous forcing differs strongly from boundary-driven systems used in most previous experimental studies on dilute granular systems, and leads to several major differences (no cluster formation, exponential tail of the particle velocity distribution) \cite{FalconEPL13}. 

Here, we study the mean dissipated power within a 3D turbulent flow homogeneously forced at small scale (particle size) via magnetic particles. We show notably that the dissipated power within the fluid increases linearly with the number $N$ of particles in the fluid, and with the magnetic field strength $B$. Beyond the agreement with the expected scalings for the magnetic injected power, we show that the dissipated power comes from two main contributions of same order of magnitude: the viscous turbulent friction of particles within the fluid, and the inelasticity of collisions, both contributions scaling linearly with $N$ and $B$ using simple scaling arguments. 
Dynamical and statistical properties of such a turbulent flow generated by this new forcing will be described elsewhere. Beyond direct interests in turbulent flows, our study also provides a model experiment in various domains such as magnetic hyperthermia for medical therapy \cite{Lesfilles}, or magnetically fluidized granular beds accelerating technological processes \cite{Rosensweig07,Penchev1990,Lupanov07}, in which particle dynamics in a fluid are controlled by an alternating magnetic field.

  \begin{figure}[t!]
    \centering
        \includegraphics[height=60mm]{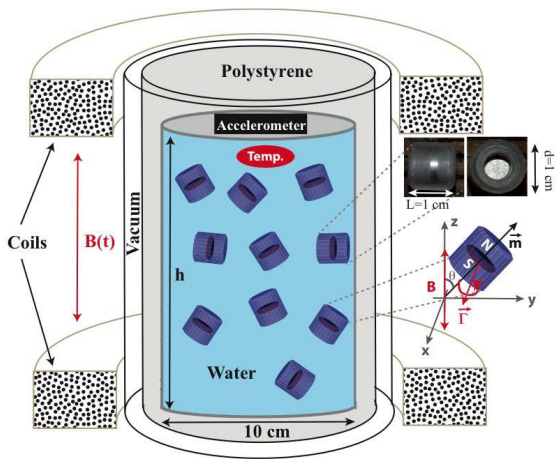}
    \caption{Experimental set up. Top insets show pictures of a magnetic particle.}
    \label{fig01}
\end{figure}

\section{Experimental setup}
The experimental setup is shown in Fig.\ \ref{fig01}. A cylindrical glass container, 10 cm in diameter and 14 cm in height, is filled with water and $N$ magnetic particles $N \in [2, 60]$ (corresponding to less than 1 layer of particles at rest). The total fluid depth $h$, including particles, is fixed in the range $h \in [4, 13]$ cm. Magnetic particles are made of a disc permanent magnet in Neodymium (NdFeB, N52) encased, and axially aligned, in a home made Plexiglas cylinder ($d$=1 cm in outer diameter, 0.25 cm in thickness, and $2R=1$ cm long) \cite{FalconEPL13}. The top of the container is closed with a Plexiglas lid in contact with water regardless $h$ (no free surface). The container is thermally insulated by inserting it within a cylindrical cell made of polystyrene foam (2.5 cm in thickness) and the latter into a vacuum flask (Dewar) -- see Fig.\ \ref{fig01}. The top flask is closed with polystyrene foam. The whole system is aligned between two coaxial coils. The coils are driven with a 50 Hz alternating current. A vertical alternating magnetic induction $B$ is thus generated in the range $0\leq B\leq 225$~G with a frequency $f=50$ Hz. $B$ is measured with a Hall probe. The Helmholtz configuration of the coils ensures a spatially homogeneous $B$ within the flask volume with a 3\% accuracy. Temperature is measured by means of thermistors (PTC Carbon) at 2 different locations. The temperature $T$ within the fluid is measured just under the Plexiglas lid. The atmosphere temperature $T_a$ of the room is measured at 40 cm far from the flask. Temporal evolutions of the temperature are recorded for $10^4$ s with a 50 Hz sampling frequency. An accelerometer attached to the top of the lid records the particle collisions with the underside of the lid for 500 s to extract the particle collision statistics with the lid. The sampling frequency was fixed at $80$ kHz to resolve collisions ($\sim 60$ $\mu$s). We focus here on the dilute regime of particles with volume fractions of $\Phi=NV_p/V < 5\%$ for temperature measurements, and of $\Phi < 14.7\%$ for collision statistics measurements, $V$ being the volume of the fluid and $V_p=\pi d^2R/2=0.78$ cm$^3$ the volume of a particle.
 
\section{Forcing mechanism}
Let $\theta$ be the angle between the magnetic moment $\vec{\mathsf{m}}$ of a particle and the vertically oscillating magnetic field $B(t)=B\sin{\omega t}$, with $\omega=2\pi f$ and fixed $f=50$ Hz (see Fig.\ \ref{fig01}). A forcing torque $\vec{\Gamma}=\vec{\mathsf{m}} \times \vec{B}$ is thus applied by the field on the magnetic moment $\vec{\mathsf{m}}$ of each particle. The fluid creates on the particle a viscous dissipative torque $\Gamma_d$ assumed at high Reynolds number (see below). The angular momentum theorem thus leads to the equation of motion of a single particle as
\begin{equation}
\ddot{\theta}(t) + \lambda \dot{\theta}(t) |\dot{\theta}(t)|=2\omega_B^2\sin{\omega t}\sin{\theta(t)}{\rm ,}\ \ \ \dot{\theta}\equiv d\theta/dt
\label{MotionEq}
\end{equation} 
with $\omega_B=\sqrt{\mathsf{m}B/(2I)}$ the resonance frequency, $\lambda= \pi \rho C_r R^5/I=0.007$ the damping coefficient with $\rho=10^3$ kg/m$^{3}$ the fluid density, $C_r=0.01$ the rotational drag coefficient of a sphere in a turbulent flow \cite{Sawatzki70,note1}, $I=1.4\ 10^{-8}$ kg\ m$^2$ the moment of inertia of the particle, $\mathsf{m}=1.55\ 10^{-2}$ A\ m$^2$ its magnetic moment, and $m=10^{-3}$ kg, its mass. Even without the dissipative term, Eq.\ (\ref{MotionEq}) is non integrable, and $\theta(t)$ then shows periodic motions, period doubling, or chaotic motions \cite{Croquette81,Meissner85}. Indeed, the oscillating magnetic field can be thought as being composed of two counterrotating fields. The equation of motion in either one of the rotating field components is integrable, and may result, depending on the initial conditions, in small oscillations of $\theta(t)$ with its own angular frequency (resonance at $\omega_B$), or in the rotation of the particle at a different frequency. This is the interaction with both field components simultaneously that renders the system nonintegrable, the second rotating field acting as  a perturbation, leading to frequency lock-ins and chaotic rotational motions \cite{Croquette81,Meissner85}. The occurrence of erratic motions is controlled by the stochasticity parameter $s\equiv 2\omega_B/\omega$ \cite{Croquette81}. 
$s$ is also linked to the ratio between the magnetic dipolar energy of the particle, $E_d=\mathsf{m}B$, and its rotation energy at the field frequency, $E_{rot}=I\omega^2/2$, since $s=\sqrt{E_d/E_{rot}}$. Periodic motions (oscillations, and uniform rotations) are predicted to occur when $s \ll 1$ (i.e. $B \ll  I\omega^2/(2\mathsf{m})=445$ G) \cite{Meissner85}. When this condition is violated, a chaotic rotational motion is predicted to occur~\cite{Croquette81}. Experimentally, we observe erratic rotational motions for $B\in [90, 225]$ G corresponding to $s \in [0.44, 0.70]$ or $\omega_B/(2\pi) \in [11, 17.4]$~Hz not so far from $\omega/(2\pi) = 50$ Hz. This lowest value of $B$ corresponds to the particle fluidization onset (see below). Particles are rotating erratically, with rotation axis mostly normal to the particle axis, frequency and direction of rotations  displaying unpredictable reversing (see \cite{FalconEPL13} for movie of particle motions in air). Numerical simulations of Eq.\ (\ref{MotionEq}) confirm these results. Indeed, for large initial conditions ($\theta_{0}$, $\dot{\theta}_0$), the particle undergoes some time dependent rotations in one or other direction as observed experimentally. The corresponding spectrum of $\dot{\theta}(t)$ displays main peaks near $\omega$ (i.e. at $\omega \pm 2\omega_B^2/\omega$) and near $\omega_B$ (i.e. at $4\omega_B^2/\omega$). For low initial conditions, oscillations of $\theta(t)$ are reported near $\omega_B$ (i.e. at $3\omega^2_B/2\omega$) superposed with ones of smaller amplitudes near $\omega$. The external magnetic field thus generates an erratic rotational driving of each particle. A spatially homogeneous forcing is thus experimentally obtained where only the rotational degrees of freedom of each particle are stochastically driven in time. Due to collisions of particles with the container boundaries or with other particles, erratic translational motions of particles are also observed. Finally, note that  numerical simulations show that the weak level of dissipation does not modify qualitatively the results obtained without dissipation. Indeed, dissipation is known to shift the onsets of lock-ins, and the value of the Feigenbaum exponent of the period doubling bifurcations towards chaos; strange attractors may also occur \cite{Meissner85, Kim99}.
Let us now give an estimate of the Reynolds number of the flow, $Re \equiv vh/\nu$, with $h$ the fluid depth, $\nu$ the fluid kinematic viscosity, and $v$ the rotational particle velocity. Assuming a synchronous particle rotation at the magnetic field frequency, i.e. $d\theta/dt=\omega$, one has $v=\omega R\simeq 1.6$ m/s, and thus typically, $Re \sim 10^5$ at most. The flow is thus turbulent as assumed by the drag torque used in Eq.\ (\ref{MotionEq}).

\begin{figure}[t!]
 \centering
        \includegraphics[height=55mm]{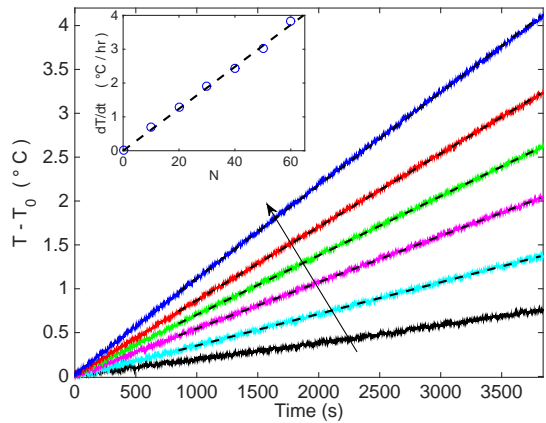} 
    \caption{Temporal evolution of the fluid temperature for different numbers of particles $N=10$, 20, 30, 40, 50 and 60 (see arrow). $B=225$ G. $h=13$ cm. Dashed lines are linear fits. Inset: fit slopes, $dT/dt$, versus $N$. Dashed line has a slope 1.7\ 10$^{-5}$ K/s.} 
    \label{fig02}
\end{figure}

\section{Thermal dissipated power in the fluid}
\label{temp}
The mean dissipated power is experimentally inferred by directly measuring the growth rate of the mean fluid temperature. Note that such a thermal measurement of the global dissipated power in turbulence flows is scarce. Indeed, to our knowledge, such direct measurements have been only performed with a spatially localized forcing at the scale of the container for hydrodynamics flows (Von K\'arm\'an or Couette-Taylor flows) \cite{Cadot1997} or for magnetohydrodynamics flows \cite{Bourgoin2002,Berhanu2007}. 
During experiments, the room temperature variation in the vicinity of the container are found to be $|\Delta T_a| \leq 0.3^{\rm o}$C for $10000$ s regardless the values of $N$ and $B$. In comparison, the increase of the fluid temperature $T-T_0$ due to the particle motions within the fluid are more than one order of magnitude larger. $T_0$ is the fluid temperature at $t=0$ when the experiment starts.

For fixed $B$, Fig.\ \ref{fig02} shows the temporal evolution of the fluid temperature $T$ for different number $N$ of magnetic particles. $T$ is found to increase linearly with time $t$, and with a growth rate, $dT/dt$, increasing with $N$. Typically, an increase up to $4^{\rm o}$C is evidenced in 1 h with 60 particles. The inset of Fig. 1 shows $dT/dt$ {\it vs.} $N$ from the slopes of each curve of the main Fig.\ \ref{fig02}. One finds that $dT/dt$ is proportional to $N$, with a proportionality constant of $\alpha=1.7\ 10^{-5}$ K/s. Note that a departure from the linear law can be observed at very long time ($> 6000$ s) due to cumulative residual thermal losses (not shown here).

\begin{figure}[t!]
 \centering
        \includegraphics[height=55mm]{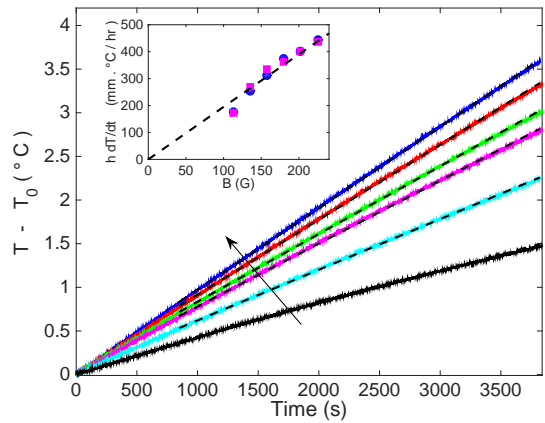} 
    \caption{Temporal evolution of the fluid temperature for different magnetic fields $B=112.5$, 135, 157.5, 180 and 202.5 G (see arrow). $N=60$. $h=13$ cm. Dashed lines are linear fits. Inset : fit slopes, $dT/dt$, versus $B$ for two fluid depths $h=6.5$ ($\bullet$) or 13 ($\blacksquare$) cm. N=60. Dashed line has a slope 5.4\ 10$^{-3}$ K/(s T).}  
    \label{fig03}
\end{figure}

Now, we fix $N$ and we increase $B$. The fluid temperature is plotted in Fig.\ \ref{fig03} as a function of time for different applied magnetic fields $B > B_c$, $B_c=90$ Gauss being the onset of particle fluidization (see \cite{FalconEPL13}). $T$ is found to increase linearly with time $t$, and with a growth rate, $dT/dt$, increasing with $B$. The inset of Fig.\ \ref{fig03} shows $dT/dt$ {\it vs.} $B$ from the slopes of each curve of the main Fig.\ \ref{fig03}. One finds that $dT/dt$ is proportional to $B$ beyond $B_c$. The thermal growth rate is thus driven by the magnetic field. Reiterating this experiment for a different value of fluid depth $h$ leads to similar results. Indeed, as shown in the inset of Fig.\ \ref{fig03}, by doubling the fluid volume, $dT/dt$ is halved regardless the value of $B$. Thus, one has $hdT/dt= \beta B$ with $\beta=5.4\ 10^{-3}$ K/(s T).  

To sum up, the thermal growth rate within the fluid is found to scale experimentally as 
\begin{equation}
\frac{dT}{dt}=\zeta \frac{NB}{V}{\rm ,\ for\ } B>B_c
\label{heatRate}
\end{equation}
with $V=Sh$ the fluid volume, $S$ the container section, and $\zeta=7.13 (\pm 0.06)\ 10^{-7}$ m$^3$ K/(s T) a dimensional constant. This accuracy on the value of $\zeta$ arises from its two estimations from $\alpha$ and from $\beta$. 
Using Eq.\ (\ref{heatRate}), the thermal dissipated power $P^{diss}_{exp}=\rho Vc_f dT/dt$ is thus found to scale experimentally as
\begin{equation}
P^{diss}_{exp}= \zeta \rho c_f NB {\rm ,\ for\ } B>B_c
\label{PdissExp}
\end{equation}
$c_f=4187$ J/(kg K) being the fluid specific heat. The volume fraction being $\Phi< 5$\%, the contribution of the solid phase to the medium specific heat is almost negligible. Thus, the thermal dissipated power within the fluid is typically 3.7 W (for $N=60$ and $B=200$ G). 

  \begin{figure}[t!]
 \centering
        \includegraphics[height=65mm]{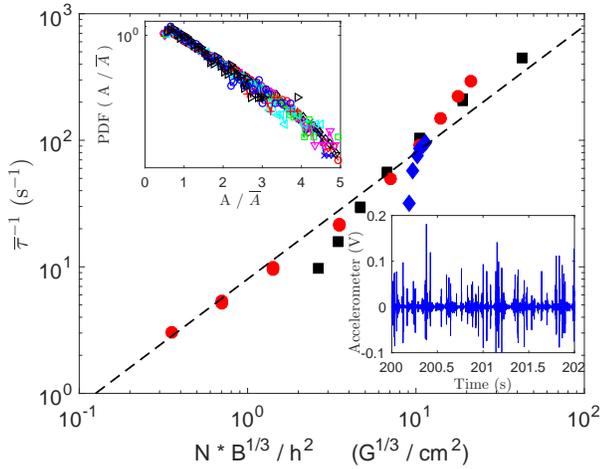} 
    \caption{Bottom inset: temporal signal of the accelerometer showing 44 collisions in 2 s ($N=10$, $h= 4$ cm, $B=180$ G). Main panel: $1/\overline{\tau}$ {\it vs.} $N B^{1/3}/h^2$ for ($\bullet$) $1\leq N \leq 60$ ($B=180$ G, $h=4$ cm), ($\blacksquare$) $2\leq h \leq 8$ cm ($N=30$, $B=180$ G), and ($\blacklozenge$) $112\leq B \leq 225$ G ($N=30$, $h=4$ cm). Top inset: PDF($A/\overline{A}$) for $1 \leq N \leq 60$ ($B=180$ G, $h=4$ cm)}  \label{fig04} 
    \end{figure}
    
\section{Collision statistics}
\label{coll}
We want now to find the scaling of the particle velocity, $v$, as a function of the parameters $N$, $B$, and $V$ as well as their collision statistics. Using an accelerometer attached to the lid, particle collisions with the lid are recorded for 500 s. A typical acceleration time series is shown in the bottom inset of Fig.\ \ref{fig04}. Each peak corresponds to the acceleration undergone by a particle during its collision on the lid. The peak amplitude, $A$, and the time lag, $\tau$, between two successive collisions on the lid are randomly distributed. A thresholding technique is applied to the signal to detect the collisions \cite{Falcon06}. 

Figure \ref{fig04} shows that the mean collision frequency scales as $1/\overline{\tau} \sim NB^{1/3}/h^2$ over 2 decades when varying one single parameter $N$, $B$ or $h$ while keeping the other two fixed. The scaling with $N$ and $h$ are the same as the ones found experimentally without liquid in the container \cite{FalconEPL13}. The different scaling found with $B$ is explained below. 

\begin{figure}[t!]
 \centering
        \includegraphics[height=65mm]{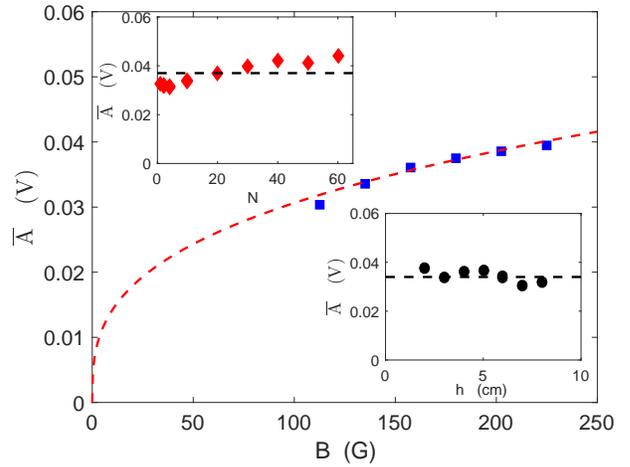}
    \caption{Mean acceleration (in Volt) during particle collisions with the top wall {\it vs.} $B$. Main panel: $\overline{A}$ {\it vs.} $B$ for $N=30$, and $h=4$ cm. Dashed line: $B^{1/3}$ scaling. Top inset: $\overline{A}$ {\it vs.} $N$ for $B=180$ G, and $h=4$ cm. Bottom inset: $\overline{A}$ {\it vs.} $h$ for $B=180$ G, and $N=30$.}  \label{fig05} 
\end{figure}

The mean amplitude of acceleration peaks is experimentally found to scale as $\overline{A} \sim h^0N^0B^{1/3}$ as shown in Fig.\ \ref{fig05}. For an impulse response of the accelerometer to a single collision, one has $|v|=A\delta t$, with $A$ the acceleration peak amplitude, $\delta t \simeq 60$ $\mu$s the roughly constant collision duration, and $v$ the magnitude of the particle velocity. Hence, the order of magnitude is $v\sim 1$ m/s, and one finds experimentally 
\begin{equation}
\overline{|v|} \sim h^0N^0B^{1/3}. 
\label{velocityExp}
\end{equation}
The particle velocity is thus independent of the number of particle $N$ and the fluid volume for our range of parameters.  Although our range of $B$ is quite small, the scaling in $B^{1/3}$ differs from the one in $B^{1/2}$ found experimentally without liquid in the container \cite{FalconEPL13}. The difference between both scalings is due to additional viscous turbulent dissipation. Indeed, the magnetic field constantly injects energy in particle rotations, the latter being converted in translation energy only during collisions without liquid. With a liquid, the viscous dissipation takes place significantly during the particle motions, and thus modifies the power budget (see below). 

As for an ideal gas, the particle velocity is found to be independent on $N$ and to be fixed only by the "thermostat", here, a magnetic one. The experimental scalings of $1/\overline{\tau}$ and $\overline{|v|}$ with $N$ and $B$ are also consistent since both quantities are usually related by the density $N/V$ as $1/\overline{\tau} \sim \overline{|v|}N/V$ \cite{Reif}. 

For various $N$ at fixed $B$, the probability density functions (PDF) of $A$ show exponential tails that collapse on a single curve when rescaled by $\overline{A}$ (see top inset of Fig.\ \ref{fig04}). Similar results are found at fixed $N$ regardless of $B$. Moreover, since $A=|v|/\delta t$, the tail of the particle velocity's PDF is thus found to scale as $\exp(- c\  v / \overline{|v|} )$ independently of the volume fraction since $\overline{|v|} \sim N^0$, $c$ being a dimensionless constant. Consistently, similar results are found for the time lag distribution. To sum up, the particle velocity distribution displays an exponential tail and is independent of the particle density. It is thus not Gaussian as for an ideal gas, or stretched exponential and density dependent as for a boundary-forced dissipative granular gas \cite{Rouyer00}. This difference with the latter case is related to the spatially homogeneous forcing used here. Indeed, a similar exponential tail independent of density has been found experimentally with this forcing without liquid in the container \cite{FalconEPL13}.

\section{Origins of the dissipated power} 
Let us first check the power budget. An estimate of the total magnetic power injected into the fluid by $N$ particles reads $P_{inj}=N\mathsf{m}\times B \omega/2 \sim NB$. These scalings in $N$ and $B$ are thus in good agreement with the ones found experimentally in Eq.\ (\ref{PdissExp}) for the thermal dissipated power $P^{diss}_{exp}$. Indeed, for an out-of-equilibrium steady state system, the mean injected power has to balance the mean dissipated power, $P_{inj}=P_{diss}$. The upper limit for $P_{inj}$, inferred from the above equation, is $P^{max}_{inj}=4.1$ W (for $N=60$, $B=200$ G), which is consistent with the experimental value of 3.7 W found above for the thermal dissipated power within the fluid.

The experimental scalings of the dissipated power with the control parameters, found in previous Sections, can be explained using simple scaling arguments. Let us first summarize the two main experimental results: the global dissipated power scales as $P^{diss}_{exp} \sim N B$ [see Eq.\ (\ref{PdissExp})] from thermal measurements, and the particle velocity $\overline{|v|} \sim N^0 B^{1/3}$ [see Eq.\ (\ref{velocityExp})] from the collision statistics. A part of the magnetically injected energy is dissipated due to the inelasticity of the collisions between particles and walls. A simple estimate of the dissipated power by the collisions is $P^{diss}_{coll}=E^{diss}_{coll}\overline{\tau}^{-1}$ with $E^{diss}_{coll}=(1-\epsilon^2)mv^2$ the dissipated energy by collisions, $\epsilon$ being the restitution coefficient, and $\overline{\tau}^{-1} \sim N B^{1/3}$ the collision frequency on a wall as found above experimentally. Using the scaling of $v$, we have thus directly $P^{diss}_{coll} \sim Nv^3 \sim NB$, which is in agreement with the experimental scalings for $P^{diss}_{exp}$. Another part of the injected energy is also dissipated by viscous friction by the turbulent drag opposing to the translational and rotational motions of particles.  On the one hand, the power dissipated by friction for translational motions reads  $P^{diss}_{trans}=NF_dv$ with $F_d\sim C_tv^2$ the turbulent drag force, and $C_t$ the translational drag coefficient. One has thus $P^{diss}_{trans}\sim Nv^3 \sim NB$, which is also in agreement with the experimental scalings for $P^{diss}_{exp}$. On the other hand, the power dissipated by viscous turbulent friction for rotational motions reads $P^{diss}_{rot}=N\dot{\theta} \Gamma_d$. One has thus $P^{diss}_{rot} \sim N\dot{\theta}^3$ which is independent of $B$ if we assume a synchronous rotation ($\omega$, i.e. $\dot{\theta}=\omega$). Consequently, $P^{diss}_{rot} \sim N\omega^3 \sim NB^0$, which differs from the experimental scalings for $P^{diss}_{exp}$. 

To sum up, from the scalings with $N$ and $B$, the magnetic power injected in rotational degrees of freedom of particles is found to be balanced by the thermal dissipated power, the latter resulting from the viscous turbulent friction of particles within the fluid on the one hand, and from the inelasticity of collisions, on the other. The order of magnitude of each dissipated power for $N=60$ is found to be of the order of 1 W (using $C_{t}=0.4$ \cite{Schlichting79}, and assuming $v=1$ m/s, $\epsilon=0.5$). Their sum is thus of the same order of magnitude than the measured dissipated power (3.7 W). It is worth noting that their ratio $P^{diss}_{trans}/P^{diss}_{coll}\sim Nv\tau \sim N^0B^0$ (using the experimental scalings of $\overline{|v|}$ and $\overline{\tau}$), and thus does not depend on the control parameters ($N$ or $B$) for our range of $Re$, but only on geometrical parameters and on properties of the fluid, particle and container.
   
\section{Conclusion}
We reported measurements of global dissipated power within a turbulent flow forced homogeneously at small scales. Unlike most laboratory experiments on 3D turbulence involving a deterministic and spatially localized forcing at large scale, the forcing used here is random in both time and space within the fluid by using magnetic particles. By measuring the growth rate of the mean fluid temperature, we show that the mean dissipated power within the fluid increases linearly with the number $N$ of magnetic particles in the system, and the strength of the magnetic field $B$. This mean energy dissipation is found to be balanced by the mean magnetically injected power, as expected for out-of-equilibrium steady state systems. Beyond this agreement, we show that the dissipated power comes from the viscous friction of particles within the fluid, and from the inelasticity of collisions, both dissipated powers scaling theoretically linearly with $N$ and $B$. By measuring the particle collision statistics on a container wall, we show that the particle velocity is independent of $N$, and is fixed only by the magnetic ``thermostat'' as $B^{1/3}$.  Finally, this new forcing mechanism appears very promising to generate fully homogeneous turbulence in order to characterize, for instance, large-scale properties of turbulence (i.e. larger than the forcing scale).

\begin{acknowledgments}
We thank M. Berhanu for providing some references. Part of this work has been supported by ANR Dysturb 17-CE30-0004-02 and ESA Topical Team N$^{\circ}$4000103461CCN2.
\end{acknowledgments}



\begin{thebibliography}{99}
\bibitem{note1}$C_r$ is roughly constant for $3\ 10^4 \lesssim {\rm Re} \lesssim 6\ 10^5$ \cite{Sawatzki70}. Its value is assumed to be the one of an effective sphere since the aspect ratio of our cylindrical particle is 1.
\end{thebibliography}
\end{document}